\documentclass{aa}
\usepackage{graphicx,epsfig,epsf}
\usepackage{longtable}

\begin{document}
\renewcommand{\topfraction}{0.85}
\renewcommand{\bottomfraction}{0.7}
\renewcommand{\textfraction}{0.15}
\renewcommand{\floatpagefraction}{0.66}

\newcommand{\hess}{H.E.S.S.}
\newcommand{\cl}{99\,\%}

\title{Search for TeV emission from the region around PSR B1706--44 with the
\hess\ experiment}

\titlerunning{Observation of PSR B1706--44 with \hess}

\author{F. Aharonian\inst{1}
 \and A.G.~Akhperjanian \inst{2}
 \and K.-M.~Aye \inst{3}
 \and A.R.~Bazer-Bachi \inst{4}
 \and M.~Beilicke \inst{5}
 \and W.~Benbow \inst{1}
 \and D.~Berge \inst{1}
 \and P.~Berghaus \inst{6} \thanks{Universit\'e Libre de 
 Bruxelles, Facult\'e des Sciences, Campus de la Plaine, CP230, Boulevard
 du Triomphe, 1050 Bruxelles, Belgium}
 \and K.~Bernl\"ohr \inst{1,7}
 \and C.~Boisson \inst{8}
% ...
}

\institute{
Max-Planck-Institut f\"ur Kernphysik, P.O. Box 103980, D 69029
Heidelberg, Germany
\and
 Yerevan Physics Institute, 2 Alikhanian Brothers St., 375036 Yerevan,
Armenia
\and
University of Durham, Department of Physics, South Road, Durham DH1 3LE,
U.K.
\and
Centre d'Etude Spatiale des Rayonnements, CNRS/UPS, 9 av. du Colonel Roche, BP
4346, F-31029 Toulouse Cedex 4, France
\and
Universit\"at Hamburg, Institut f\"ur Experimentalphysik, Luruper Chaussee
149, D 22761 Hamburg, Germany
\and
Physique Corpusculaire et Cosmologie, IN2P3/CNRS, Coll{\`e}ge de France, 11 Place
Marcelin Berthelot, F-75231 Paris Cedex 05, France
\and
Institut f\"ur Physik, Humboldt-Universit\"at zu Berlin, Newtonstr. 15,
D 12489 Berlin, Germany
\and
LUTH, UMR 8102 du CNRS, Observatoire de Paris, Section de Meudon, F-92195 Meudon Cedex,
France
% ...
}

\offprints{Bruno.Khelifi@mpi-hd.mpg.de}

\date{Accepted by Astronomy \& Astrophysics}

\abstract{

The region around PSR~B1706--44 has been observed with the \hess\ imaging atmospheric Cherenkov telescopes in
2003. No evidence for $\gamma$-ray emission in the TeV range was found at the pulsar position or at the radio arc
which corresponds to the supernova remnant G\,343.1--2.3. The \cl\ confidence level flux upper limit at the
pulsar position is F$_{\textrm{ul}}($E$>$$350\,\textrm{GeV}) = 1.4 \times 10^{-12} \, \textrm{s}^{-1}
\textrm{cm}^{-2}$  assuming a power law ($dN/dE \propto E^{-\Gamma}$) with photon index of $\Gamma$$=$$2.5$ and
F$_{\textrm{ul}}($E$>$$500\,\textrm{GeV}) = 1.3 \times 10^{-12} \, \textrm{s}^{-1} \textrm{cm}^{-2}$ without an
assumption on the spectral shape. The reported upper limits correspond to 8\,\% of the flux from an  earlier
detection by the CANGAROO experiment.

}

\maketitle

\section{Introduction}

PSR\,B1706--44 is a young pulsar (spin-down age of $\sim$17\,kyrs) with distance estimates ranging from 1.8 to
3.2\,kpc with a period of 102\,ms and a spin-down luminosity of about 1\% of the Crab pulsar ($3.4\times
10^{36}\,\textrm{erg}\,\textrm{s}^{-1}$). Pulsed emission has been observed at radio and X-ray  wavelengths, and
in GeV $\gamma$-rays. An extended synchrotron nebula around this compact object has been found in radio 
observations (\cite{vla}) with an extension of 1'--4' and with a flat spectrum (energy index of $0.3$), and
also in X-rays (\cite{chandra}) with an extension of $\sim$$20$" and with a photon index of $1.34$. These
characteristics suggest the existence of a pulsar wind nebula (PWN) powered by the pulsar. In the TeV range, the
CANGAROO experiment detected a steady emission  coincident with the PWN position at a level of roughly 50\,\% of
the Crab flux (Kifune et al. \cite{cangaroo}; \cite{cangaroo_icrc}),  suggesting that this PWN is the southern
equivalent of the Crab nebula.  The Durham~Mark~6 collaboration (Chadwick et al. \cite{durham}) reported also a
significant detection above 300\,GeV. A flux upper limit above $500$\,GeV which is compatible with  the CANGAROO
flux has been derived using data from the BIGRAT telescope (Rowell et al. \cite{gavin}).

PSR\,B1706--44 is coincident with an incomplete arc of radio emission (\cite{mcadam}) which has been interpreted
as a shell-type supernova remnant (SNR) named G\,343.1--2.3. This SNR has been detected only at radio wavelengths
(\cite{duncan}) and may be associated with the pulsar as discussed in \cite{bock}.

We present here the results of the observation of the field of view around PSR~B1706--44 with the \hess\
experiment.  \hess\ is an atmospheric Cherenkov detector dedicated to the observation of TeV $\gamma$-rays
(\cite{HESS}). Situated in Namibia, the full four-telescope array is operational since December 2003. Each
telescope has a mirror area of 107~$\textrm{m}^2$ (\cite{optics}) and is equipped with a camera consisting of
960~photomultiplier tubes (PMT) (\cite{camera}). The system has a field of view of $5^\circ$. In stereoscopic
observation mode, it allows one to reconstruct the  direction of individual showers with a precision better than
$0.1^\circ$. 

\section{Observations and data analysis}

PSR~B1706--44 was observed with two \hess\ telescopes between April and July 2003. During this commissioning phase,
GPS time stamps were used in the offline data analysis to identify showers observed in coincidence by the two
telescopes. This coincidence requirement allows for a higher background rejection and thus for a better sensitivity
than single telescope observations. In this configuration, a source with a flux of 5\% of Crab nebula can be detected
with more than $5\,\sigma$ in 4.5\,hours at $20^\circ$ zenith angle. The pulsar was observed with 28-min runs in
wobble mode, whereby runs are taken pointing $\pm 0.5^{\circ}$ away from the pulsar position in declination. Data
affected by hardware problems or bad weather conditions were excluded from analysis. The proper functioning of the
detector system was verified by numerous checks. The telescope pointing has been confirmed by  correlating high PMT
currents with bright stars in the field of view.  The trigger rate of the system is well reproduced by simulations for
cosmic rays, and the shape of simulated $\gamma$-ray images is consistent with the result of Crab observations. Data
analysis is performed with two completely independent chains with different calibrations, with independent Monte Carlo
simulations and with different analysis techniques.

\begin{figure}
\centering
\includegraphics[width=9.5cm]{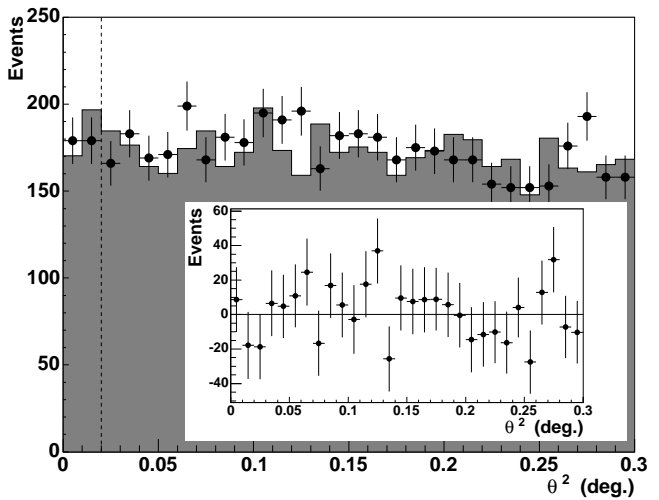}
\vspace{-0.6cm}
\caption{$\theta^2$ distribution calculated with respect to the PWN position. The dots denote events from the ON region,
the histogram are the events from the OFF region scaled by the normalization factor $\alpha$. The dashed vertical line
indicates the applied angular cut. The inset shows the difference between the ON and the scaled OFF regions.}
\label{fig_theta}
\vspace{-0.3cm}
\end{figure}

The selected data have a total live time of $14.3$ hours. The energy threshold estimated from Monte Carlo simulations
at the average observation zenith angle ($\sim$$26^{\circ}$) is about 350\,GeV. This threshold is higher than for the
four-telescope system since the telescopes were operated with higher trigger thresholds  in the commissioning phase.
Data were analysed using standard shower reconstruction and standard background rejection methods (\cite{2155}).
Standard cuts, optimised on Monte Carlo simulations, have been applied on mean scaled Hillas parameters in order to
increase the signal-to-background ratio. Showers were classified using the angular distance $\theta$ between their
reconstructed direction and the direction of possible source. For this standard analysis, showers were accepted as
coming from the source (the ON region) when their $\theta^2$ was smaller than $0.02\,\textrm{degree}^2$ (i.e.~angular
distance smaller than $8.5'$). The background was determined by counting events in a ring (the OFF region) centered at
the investigated direction whose inner radius is larger ($>$$0.4^\circ$) than the $\theta^2$ cut and whose area is 7
times larger than the ON region. A normalization factor $\alpha$ is applied to these estimated background counts to
correct for the different size of ON and OFF regions and the different radial acceptance in the field of view.

\section{Results}

\begin{figure}[t!]
\vspace{-0.4cm}
\centering
\includegraphics[width=8.5cm]{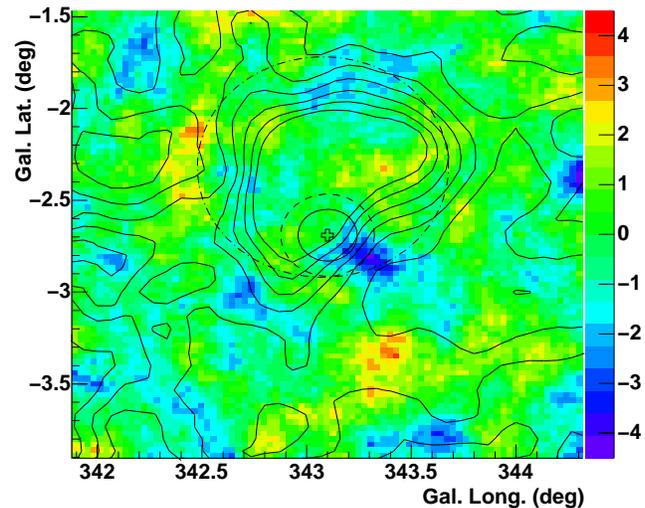}
\vspace{-0.3cm}
\caption{Significance map centered on PSR~B1706--44. The cross marks the
pulsar position. The contour lines correspond to the 2.2\,GHz image of G\,343--2.3 (\cite{duncan}). The solid
circle indicates the integration region of the {\em Standard} cuts, the dashed circle the
{\em CANGAROO} cuts and the dot-dashed circle the {\em Radio arc} cuts. The significance distribution for the
entire \hess\ field of view is compatible with a Gaussian of mean $-0.06$ and of sigma $1.09$.}
\label{fig_map}
\end{figure}

A plot of $\theta^2$ relative to the PWN position is shown in Figure~\ref{fig_theta}. The significance, 
calculated according to \cite{lima}, is $0.1\,\sigma$. Table~\ref{tab_results} provides an overview of the event
statistics in the column labelled {\em Standard}. The analysis described above was repeated with the same cuts
for every point in the field of view. The resulting significance map is presented in Fig.~\ref{fig_map}. It
exhibits no significant point source excess in the vicinity of the pulsar or on the radio emission arc.

In order to roughly reproduce the conditions of the PSR~B1706--44 detection by CANGAROO, the analysis at the pulsar
position was repeated using a looser $\theta^2$ cut of  $0.05\,\textrm{degree}^2$ and selecting events above an energy
of 1\,TeV. The results are shown in the column labelled {\em CANGAROO} of Table~\ref{tab_results}  and give no
indication for a significant excess. For the analysis of the radio arc, a $\theta^2$ cut of  $0.36\,\textrm{degree}^2$
has been applied around the position (17h08m,$-44^\circ$17') and no significant excess is measured (column labelled
{\em Radio arc} of Table~\ref{tab_results}).

\vspace{-0.6cm}
\begin{table}[h!]
 $$ 
       \begin{tabular}{c c c c}
         \hline
         \noalign{\smallskip}
                            &{\em Standard} &{\em CANGAROO}   &{\em Radio arc}\\
         \noalign{\smallskip}
         \hline
         \noalign{\smallskip}
          $N_\mathrm{on}$   &352            &112              &4746\\
          $N_\mathrm{off}$  &2243           &512              &13688\\
          $\alpha$          &0.15620        &0.19258          &0.34592\\
          Excess            &$1.6\pm20.2$   &$13.4\pm11.1$    &$11.0\pm79.9$\\
          Significance      &$0.1\,\sigma$  &$1.2\,\sigma$    &$0.1\,\sigma$\\
        \noalign{\smallskip}
         \hline
      \end{tabular}
  $$ 
\vspace{-0.3cm}
   \caption[]{Analysis results: $N_\mathrm{on}$ and $N_\mathrm{off}$ are the event numbers in the ON and 
   OFF regions, $\alpha$ is the normalisation factor. The results are reported for the
   standard $\theta^2$ cut (column labelled {\em Standard}), for the cuts reproducing the
    conditions of the CANGAROO detection (column {\em CANGAROO}) and for the analysis of the radio arc (column 
    {\em Radio arc}).}
   \label{tab_results}
\vspace{-0.6cm}
\end{table}

Limits on the integral flux above certain energies $E_T$ were obtained using two different methods. The first
method (Method~A) tests the hypothesis that the number of excess events with energies above $E_T$ result from a
source with a power law spectrum  with a (positive) photon index $\Gamma$. The photon index was varied between $2$
and $3$. This range includes the value of $\Gamma$$=$$2.5$ from the  earlier CANGAROO detection. The second method
(Method~B) makes no assumption about the source spectrum and calculates the integrated flux $F$ directly as the
difference of the measured flux from the ON region and the flux of cosmic-rays from the OFF region:

\[
        F(>E_T) = \frac{1}{T}\left(
        \sum_{i=1}^{N_{\mathrm{on }}} \frac{1}{A_i} - \alpha                                                 
        \sum_{i=1}^{N_{\mathrm{off}}} \frac{1}{A_i} \right) \mbox{.}
 \label{eq:one}
\]

Here, $T$ is the live time, and both sums on the ON and OFF regions run over all showers with reconstructed
energies greater than $E_T$. The effective areas ($A_i$) depend on the zenith angle and energy of
each event, and $\alpha$ is the normalization factor. As $A_i$ is determined using the reconstructed energy, the
energy threshold should be increased such that the bias of the reconstructed energy is less than 10\,\%.  The
upper limits derived with both methods were calculated using the unified approach of Feldman \& Cousins
(\cite{new}) and a confidence level of \cl. To compare the upper limits from Method~B with a prediction, the
investigated model spectrum must be integrated over all energies starting at $E_T$.

Table~\ref{table_ul} gives the values of flux upper limits at \cl\ confidence level for various cuts and methods;
both methods give similar results. With method A, the upper limit at the PWN position corresponds to $\sim$$1$\%
of the flux from the Crab Nebula (at the same energy threshold) and the upper limit for the radio arc corresponds
to $\sim$$5$\% of the flux from the Crab Nebula. The upper limit which reproduces the experimental
conditions of the CANGAROO experiment corresponds to $\sim$$8$\,\% of the flux reported by that collaboration.

\begin{figure}
\vspace{-0.4cm}
\centering
\includegraphics[width=9.cm]{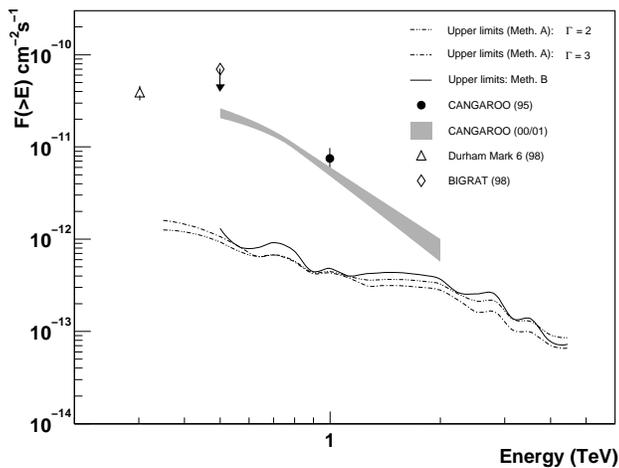}
\vspace{-0.6cm}
\caption{Integral upper limits at \cl\ CL for the flux from the PWN position (solid, dotted and dot-dashed line).
The filled circle corresponds to the CANGAROO detection of Kifune et al. (\cite{cangaroo}) and the CANGAROO integrated flux (grey
area) is calculated from the result of a broken power law fit to the 2000 and 2001 differential spectrum
(\cite{cangaroo_icrc}). The open diamond and the triangle are from Rowell et al. (\cite{gavin}) and Chadwick et
al. (\cite{durham}), respectively.}
\label{fig_ul}
\vspace{-0.3cm}
\end{figure}

\vspace{-0.5cm}
\begin{table}[h!]
 $$ 
       \begin{tabular}{c c c c c}
         \hline
         \noalign{\smallskip}
         &\multicolumn{2}{c}{{Method~A}}  &\multicolumn{2}{c}{{Method~B}}\\
         \noalign{\smallskip}
         \hline
         \noalign{\smallskip}
          Standard &$1.4\times 10^{-12}$ &($0.35$)        &$1.3\times 10^{-12}$ &($0.50$) \\
          CANGAROO &$6.4\times 10^{-13}$ &($1.00$)        &$7.7\times 10^{-13}$ &($1.00$)\\
          Radio arc &$5.8\times 10^{-12}$ &($0.35$)        &$3.5\times 10^{-12}$ &($0.50$)\\
        \noalign{\smallskip}
         \hline
      \end{tabular}
  $$ 
\vspace{-0.3cm}
   \caption[]{Flux upper limits at \cl\ confidence level for the pulsar position in $\textrm{s}^{-1} \textrm{cm}^{-2}$. 
   The upper limits from Method~A were calculated assuming a photon index of $\Gamma$$=$$2.5$. The numbers
   in parentheses are the energy thresholds (in TeV) for which the upper limits were determined. }
   \label{table_ul}
\vspace{-0.6cm}
\end{table}

\section{Discussion and Conclusions}

The reported upper limits on the flux of TeV $\gamma$-rays are roughly one order of magnitude lower than the reported
CANGAROO flux and a factor of 55 lower than earlier limits (Rowell et al. \cite{gavin}). The CANGAROO observations
were not contemporaneous with the \hess\ observations, which raises the question of whether the TeV emission could be
variable on a time scale of years. Such a variability seems unlikely given our current understanding of PWN
(\cite{pwn}). Another potential reason for the discrepancy  could be an object confusion along the line of sight.
There are, however, no BL~Lac objects or variable galactic TeV $\gamma$-ray emitters known around the pulsar. It has
been pointed out (Aharonian et al. \cite{Felix}; \cite{cangaroo_icrc}) that the high flux level reported by CANGAROO is
surprising. Since the X-ray luminosity is about 0.01\,\% of that of the Crab PWN,  the TeV $\gamma$-rays should be
emitted from a much larger volume than the X-rays,  according to the inverse Compton (IC) scenario.

Using the \hess\ flux upper limit above $1$\,TeV, a lower limit on the magnetic field can be derived from Eq.~6 of
Aharonian et al. (\cite{Felix}). This requires a measurement of the flux in the X-ray band from the same electron
population  that emits the hypothetical TeV radiation. Measurements by Chandra (\cite{chandra}) provide a flux
from the PWN, excluding the point-like emission of the central source; however, their chosen analysis region
(radius less than $10''$) is smaller than the full extent of the PWN, for which Finley et al. (\cite{finley})
found a best-fit exponential scale length of $27''$. The flux measured by ASCA (Finley et al. \cite{finley})
encompasses the entire PWN, but also includes the pulsar emission. To estimate the PWN flux, we used the ASCA
spectrum but subtracted a point source contribution estimated from ROSAT HRI to be $(43 \pm 12)$\% (Finley et al.
\cite{finley}), yielding an unabsorbed flux of $5.5 \times 10^{-13}$\,erg s$^{-1}$ cm$^{-2}$ in the 2--10\,keV
band for the PWN. The lack of observable X-ray emission below about 0.5\,keV due to interstellar absorption means
that the electrons producing the observed X-rays have somewhat higher energy than those producing TeV
$\gamma$-rays, and an extrapolation of the X-ray spectrum to lower energies is necessary. The spectral index
measured with ASCA, $\Gamma$$=$$1.7_{-0.4}^{+0.5}$, is fully compatible with the more precise determination from
BeppoSAX, $\Gamma$$=$$1.69 \pm 0.29$ (\cite{mineo}). The derived lower limit on the magnetic field strength is
then about 1\,$\mu$G when one assumes that the inverse Compton scattering involves only the photons of the
microwave background radiation and assuming the same photon index in the X-ray and TeV band. This value is however
not very constraining given that the mean Galactic magnetic field is of the same order of magnitude.

\begin{acknowledgements}

The support of the Namibian authorities and of the University of Namibia in facilitating the construction and
operation of H.E.S.S. is gratefully acknowledged, as is the support by the German Ministry for Education and
Research (BMBF), the Max Planck Society, the French Ministry for Research, the CNRS-IN2P3 and the Astroparticle
Interdisciplinary Programme of the CNRS, the U.K. Particle Physics and Astronomy Research Council (PPARC), the
IPNP of the Charles University, the South African Department of Science and Technology and National Research
Foundation, and by the University of Namibia. We appreciate the excellent work of the technical support staff in
Berlin, Durham, Hamburg, Heidelberg, Palaiseau, Paris, Saclay, and in Namibia in the construction and operation
of the equipment. We would also like to thank the Australia Telescope National Facility (ATNF) for provision of
2.2\,GHz radio data.

\end{acknowledgements}


\begin{thebibliography}{}
%
\bibitem[1997]{Felix} Aharonian, F., Atoyan A.M. and Kifune T.,
1997, \mnras, 291, 162
%
\bibitem[Aharonian et al. 2004]{2155} Aharonian, F., Akhperjanian, A.G., Aye, K.-M., et al.,
2004, accepted by Astron. \& Astrophys. (astro-ph/0411582)
%
\bibitem[Becker et al. 1995]{rosat} Becker, W., Brazier, K.T.S., Tr\"umper, J.,
1995, Astron. \& Astrophys., 298, 528	
%
\bibitem[Bernl\"ohr et al. 2003]{optics} Bernl\"ohr, K., Carrol, O., Cornils, R., et al.,
2003, APh., 20, 111
%
\bibitem[Blondin et al. 2001]{pwn} Blondin, J.M., Chevalier, R.A., Frierson, D.M.,
2001, \apj, 563, 80	
%
\bibitem[Bock \& Gvaramadze (2002)]{bock} Bock, D.C.-J. \& Gvaramadze, V.V.,
2002, Astron. \& Astrophys., 394, 533	
%
\bibitem[1998]{durham} Chadwick, P.M., Dickinson, M.R., Dipper, N.A., et al.,
1998, APh, 9, 131
%
\bibitem[Duncan et al. 1995]{duncan} Duncan, A.R., Steward, R.T., Haynes, R.F., Jones, K.L.,
1995, \mnras, 277, 36
%
\bibitem[1998]{new} Feldman, G.J. \& Cousins, R.D.,
1998, \prd, 57, 7
%
\bibitem[1998]{finley} Finley, J.P., Srinivasan, R., Saito, Y., et al.,
1998, \apj, 493, 884
%
\bibitem[Giacani et al. 2002]{vla} Giacani, E.B., Frail, D.A., Goss, W.M., Vieytes, M.,
2001, \aj, 121, 313
%
\bibitem[Gotthelf et al. 2002]{chandra} Gotthelf, E.V., Halpern, J.P., Dodson, R.,
2002, \apj, 567, L125
%
\bibitem[Hofmann 2003]{HESS} Hofmann, W., 2003,
Proc. 28th ICRC, Tsukuba, Univ. Academy Press, Tokyo, p. 2811
%
\bibitem[1995]{cangaroo} Kifune, T., Tanimori, T., Ogio, S., et al.,
1995, \apj, 438, L91
%
\bibitem[Kushida et al. 2003]{cangaroo_icrc} Kushida, J., Tanimori, T., Kubo, H., et al.,
Proc. 28th Int. Cosmic Ray Conf., Tsukuba (2003), Univ. Academy Press, Tokyo, p. 2493 
%
\bibitem[Li \& Ma (1983)]{lima} Li, T.-P. \& Ma, Y.-Q.,
1983, \apj, 272, 317
%
\bibitem[McAdam et al. 1993]{mcadam} McAdam, W.B., Osborne, J.L., Parkinson, M.L.,
1993, Nature, 361, L516
%
\bibitem[Mineo et al. 2002]{mineo} Mineo, T., Massaro, E., Cusumano, G.,Becker, W.,
2002, Astron. \& Astrophys., 392, 181
%
\bibitem[1998]{gavin} Rowell, G.P., Dazeley, S.A., Edwards, P.G., Patterson, J.R., Thornton, G.J., 
1998, APh, 194, 332
%
\bibitem[Vincent et al. 2003]{camera} Vincent, P., Denance, J.-P., Huppert, J.-F., et al. ,
2003, Proc. 28th ICRC, Tsukuba, Univ. Academy Press, Tokyo, p. 2887
%
\end{thebibliography}
\end{document}